\documentclass[aps,prl,showpacs,twocolumn]{revtex4}
\usepackage{graphicx}
\usepackage{times}

\begin{document}

\title{Regular atomic narrowing of Ni, Fe and V nanowires resolved by 2D correlation analysis}

\author{A.~Halbritter$^1$, P.~Makk$^1$, Sz.~Mackowiak$^{2,4}$, Sz.~Csonka$^1$, M. Wawrzyniak$^3$, J.~Martinek$^2$}
\affiliation{$^1$Department of Physics, Budapest University of
Technology and Economics and \\
Condensed Matter Research Group of the Hungarian Academy of Sciences, 1111 Budapest, Budafoki ut 8., Hungary}
\affiliation{$^2$Institute of Molecular Physics, Polish Academy of Sciences, 60-179 Pozna\'n, Poland}
\affiliation{$^3$Faculty of Electronics and Telecommunications, Pozna\'n University of Technology, Pozna\'n, Poland}
\affiliation{$^4$Institute of Physics, Pozna\'n University of Technology, 60-965
Pozna\'n, Poland}

\date{\today}

\begin{abstract}
We present a novel statistical method for the study of stable atomic configurations in breaking nanowires based on
the 2D cross-correlation analysis of conductance versus electrode separation traces. Applying this method, we can clearly resolve the typical evolutions of the conductance staircase in some transition metal nanojunctions (Ni, Fe, V) up to high conductance values. In these metals our analysis demonstrates a very well ordered atomic narrowing of the nanowire, indicating a very regular, stepwise decrease of the number of atoms in the minimal cross section of the junction, in contrast to the majority of the metals. All these features are hidden in traditional conductance histograms.
\end{abstract}

\maketitle

The study of electronic transport on the single atom or single-molecule scale have revealed several interesting phenomena due to the interplay of the atomic granularity of matter and the quantum
nature of conductance \cite{Agrait_review}. The conductance of stable, frequently occurring atomic configurations or
molecular contacts
formed during the rupture of metallic nanowires can be identified using peaks in the conductance histograms constructed from
a large number of conductance versus electrode separation traces \cite{Agrait_review,Yanson_Auworkhard,Yanson_Alworkhard,Csonka_AuH2,Lortscher_benzenedithiol,Gonzalez_statistics,Venkataraman_switch}.
Individual conductance traces exhibit several plateaus up to tens of the conductance quantum unit, $G_0=2e^2/h$ in accordance with simulations predicting a stepwise decrease of the number of atoms in the minimal cross-section (MCS) of the junction \cite{Pauly_PtNi,Mochales_Ni}. In the most well studied system, gold the discreteness of the MCS can be reflected by numerous peaks in the histogram \cite{Yanson_Auworkhard}. However, in the majority of the cases the histograms only show a single or a few peaks \cite{Agrait_review}; thus the knowledge about the details of contact rupture is very limited. This is especially true for transition metal nanojunctions, where the conduction through low-symmetric $d$ orbitals introduces an extreme sensitivity to the contact geometry
giving rise to a broad variety of conductances for contacts with the same MCS \cite{Jacob,Pauly_PtNi}.

For a deeper understanding of precise atomic arrangements in the junction the conductance histograms must be supplemented with further experimental input, as well as by detailed theoretical ab-initio simulations.
It was already recognized that additional statistical analysis of the conductance traces can provide fundamental information, like the discovery of monoatomic chains by plateaus' length analysis \cite{Yanson_chain}, or the deeper understanding of single-molecule junctions by recently developed statistical methods \cite{Csonka_AuH2,Lortscher_benzenedithiol,Gonzalez_statistics,Venkataraman_switch}.

In this Letter we introduce a novel characterization method based on the 2D cross-correlation histogram (2DCH) analysis of conductance traces \cite{Wawrzyniak09} in some analogy to 2D correlation spectroscopy (COSY) in magnetic resonances \cite{Ernst00}.
In the field of mesoscopic physics it is already known that correlation analysis provides significantly new information compared to the measurement of mean values, which was demonstrated by the success of quantum noise studies \cite{Nazarov03}. Implementing these ideas we demonstrate that the statistical cross-correlation analysis of conductance traces opens a possibility to answer several questions concerning the atomic-scale nanocontact formation dynamics.
Applying the correlation technique we study the contact evolution in some transition metal nanojunctions (Ni, Fe, V), demonstrating that the conductances of plateaus related to different atomic configurations show  correlated fluctuations. In the conductance histogram these fluctuations introduce a strong smearing; however, the 2DCH clearly resolves the alternative evolutions of contact conductance demonstrating a very regular stepwise change of the MCS starting from rather high conductance values. In comparison, this very regular narrowing of the nanowire is not detected in the majority of the metals.

\begin{figure}
\centering
\includegraphics[width=\columnwidth]{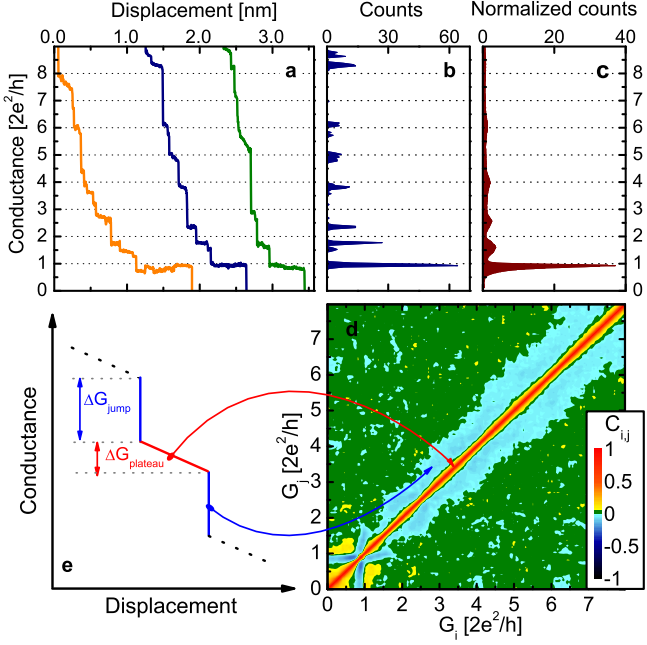}
\caption{(Color online) Measurements on Au junctions: (a) representative conductance vs electrode separation traces; (b)
histogram for the middle trace on panel (a); (c) histogram for the whole data set including 5000 traces; (d) 2D
correlation histogram, $C_{i,j}$ for
the same data set. (e) Illustration for slant
plateaus and neighboring jumps respectively being responsible
for positively correlated areas at the diagonal of the correlation
plot and the neighboring negatively correlated region.
}
\label{fig1.fig}
\end{figure}

\paragraph{2D correlation histograms.---}
We begin the discussions using our result for Au nanowires chosen as a well characterized reference system.
The measurements were performed on high purity samples applying the mechanically controllable break junction technique at liquid helium temperature \cite{supplementary}.
Typical conductance vs electrode separation traces feature some similarities in the general character and differences in the details due to the stochastic nature of contact rupture [Fig.~\ref{fig1.fig}(a)].
A conductance histogram for a chosen trace, $r$ can be constructed by counting the number of data points in discrete bins, $i$ along the conductance axis: $N_i(r)$ [Fig.~\ref{fig1.fig}(b)]. The histogram for the whole data set including several thousand traces [Fig.~\ref{fig1.fig}(c)] is defined by the average number of data points in the different bins: $H_i=\langle N_i(r) \rangle_r$.

Though conductance histograms have a fundamental role in the study of atomic and molecular nanojunctions, it is evident that histograms grab only partial knowledge from the measured data. Significantly more information can be collected by  studying the
cross-correlations between different atomic configurations.
The cross-correlation function between bin $i$ and $j$ can be defined as:
\begin{equation}
 \label{eq1.eq}
    C_{i,j}=\frac{\left\langle \delta N_i(r) \cdot \delta N_j(r) \right\rangle_r}
    {\sqrt{\left\langle [\delta N_i(r)]^2\right\rangle_r\left\langle [ \delta N_j(r) ]^2 \right\rangle_r}} \;,
    \end{equation}
where $ \delta N_i(r) \equiv N_i(r)-\langle N_i(r) \rangle_r $.
This function gives zero if $N_i(r)$ and $N_j(r)$ are independent from each other. If the occurrence of two plateaus near bins $i$ and $j$ is correlated -- i.e. either both of them or none of them appear -- the correlation function takes positive values. Contrary, it takes negative values if two plateaus are anticorrelated; i.e., the occurrence of a plateau near bin $i$ excludes the occurrence of another one at bin $j$.
By definition the correlation is limited to the range $-1\le C_{i,j}\le 1$, it is symmetric, $C_{i,j}=C_{j,i}$, and the diagonal always gives  $C_{i,i}=1$. The cross-correlations can be visualized by a 2D color plot, where the two axis correspond to bins $i$ and $j$ and the color-scale shows $C_{i,j}$ with warm colors (yellow-red) for positive correlations, cold colors (blue-black) for negative correlations, and the areas where the correlation is zero within the accuracy of the method is marked by green.

In Fig.~\ref{fig1.fig}(d) we present the 2DCH for Au nanojunctions.
An evident feature observed on the 2DCH is a finite width positively correlated region near the diagonal, and a wider negatively correlated region near both sides of the diagonal. This feature has a simple reason: if on a certain trace a plateau is observed near bin $i$ (i.e. the number of points in bin $i$ is larger than in average) then, assuming a plateau with finite slope, also the neighboring bins will have larger number of points than in average, introducing positive correlation. Further away from bin $i$ jumps are observed on the trace, introducing negative correlation [see demonstration in Fig.~\ref{fig1.fig}(e)]. For even larger conductance differences no significant correlations are detected.

\begin{figure}
\centering
\includegraphics[width=\columnwidth]{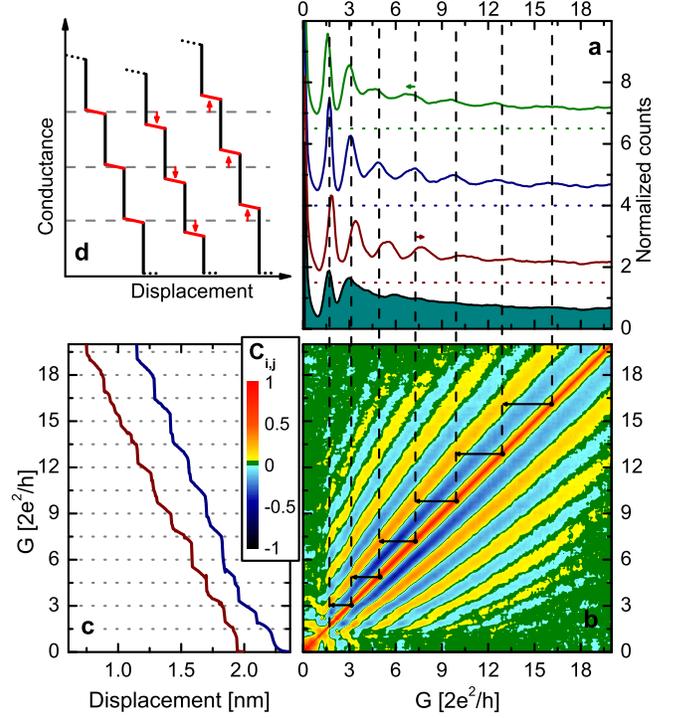}
\caption{(Color online) Measurements on Ni junctions. (a): Bottom area graph shows the conductance histogram for the whole
data set including 5000 traces. The line graphs show conditional histograms
for different subsets of the traces (see text).
Horizontal dashed lines show the zero values of the histograms, vertical dashed lines indicate the peak positions in the middle conditional histogram.
(b) 2D correlation histogram for the whole data set.
(c) Representative conductance traces. (d) Illustration for the correlated shifting of plateaus.}
\label{fig2.fig}
\end{figure}

\paragraph{Correlation analysis of Ni, Fe and V junctions.---}
We have performed 2DCH analysis of some transition metal nanojunctions: Ni, Fe, V. Here, we present and discuss our results on Ni junctions in detail [Fig.~\ref{fig2.fig}], while the experiments on Fe and V exhibiting similar features are demonstrated in the supplementary material \cite{supplementary}, and discussed in detail elsewhere \cite{Makk}.
The typical histogram of Ni, Fig.~\ref{fig2.fig}(a) exhibits two peaks at $G=1.7 \, G_0$ and $G=3 \, G_0$, but at higher conductances no clear features are resolved, in accordance with previous low-temperature studies  \cite{Undtiedt_Kondo,Untiedt_NiCoFe}. In contrast, individual conductance traces [Fig.~\ref{fig2.fig}(c)] show clear plateaus up to tens of the conductance quantum unit. The conductance of the plateaus exhibit considerable variation from trace to trace which causes a complete smearing of the histogram at higher conductance values, and only two low-conductance peaks survive.

We present the 2DCH for Ni contacts in Fig.~\ref{fig2.fig}(b), where
similarly to Au junctions, an
anticorrelated region is seen beside the diagonal, however further away several strong positively correlated {\it stripes} are observed, which are almost parallel with the diagonal.
The regions positively correlated at the coordinate $(i,j)$ correspond to
the simultaneous appearance of plateaus both at bins $i$ and $j$; thus the first stripe close to diagonal
defines the probable conductance pairs of neighbor plateaus.
The presence of stripes indicates that the conductances of plateaus change in a correlated way: if the conductance of a certain plateau shifts upwards or downwards from trace to trace due to the changes in the contact environment, the conductances of neighbor plateaus are shifted in correlated way leaving the differences in conductance similar [see demonstrative cartoon in Fig.~\ref{fig2.fig}(d)].

To better resolve the nature of the correlation, in Fig.~\ref{fig2.fig}(a) we plot separate histograms for selected traces that have a plateau at a certain conductance - the top, middle and bottom curves correspond to conductance traces which have the final plateau near $G = 1.6, 1.7, 1.8 \, G_0$, respectively.
Precisely, these so called {\it conditional histograms} are constructed from those traces, $r'$, for which the number of data points in a certain bin, $j$ is larger than its average value: $H_{i,j}^{cond} = \langle N_i(r')|\delta N_j(r')>0\rangle_{r'}$. With this objective selection criterion the traces having a long enough plateau near bin $j$ can be sorted out with reasonable certainty. These conditional histograms exhibit up to several
peaks, which is the most pronounced for the middle graph, where the final plateau before complete rupture is fixed at the position of the first peak in the total histogram.
For traces with somewhat smaller or larger conductance of the last plateau, all the peaks are respectively shifted to lower or higher values. This demonstrates that the nearly featureless total histogram builds up from partial histograms with several well-defined peaks.

On the magnetic samples, Ni \& Fe control measurements were performed in high magnetic field.
A magnetic origin of the reported features can be excluded by observing similar conductance histograms and 2DCHs at $B=12\,$T.

\begin{figure}
\centering
\includegraphics[width=0.8\columnwidth]{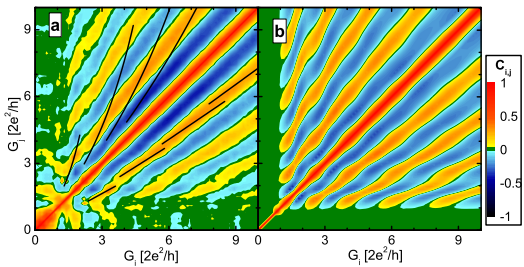}
\caption{(Color online) Simulations for stripes. (a) Experimental data on Ni with two simplified fits. (b) More advanced simulations. (See text and supplementary material for details).}
\label{fig3.fig}
\end{figure}

\paragraph{Correlated shifting of plateaus.---}
Both the 2DCH and the conditional histograms for Ni nanojunctions clearly show the correlated shifting of plateaus from trace to trace. For a better understanding of this phenomenon we have performed simple simulations for the observed stripes, which are demonstrated in Fig.~\ref{fig3.fig}, and discussed in more detail in the supplementary material \cite{supplementary}. The resistances of plateaus ($R=G^{-1}$) can be modeled as the resistance of some {\it base} configurations being modified due to the random variations in the contact neighborhood, $R_{n,r}^{\text{total}}=R_{n}^{\text{base}}+\Delta R_{n,r}$, where $n$ and $r$ label the different plateaus and traces, respectively. In Fig.~\ref{fig3.fig}(a) we present the results of two simple models, in which the variation of the resistance is random for the different traces, but on a certain trace it is either constant for all the configurations ($\Delta R_{n,r}=\Delta R_{r}$, curves above diagonal), or it is proportional to the base resistance ($\Delta R_{n,r}=\alpha_r\cdot R_{n}^{\text{base}}$, curves below diagonal). Fig.~\ref{fig3.fig}(b) shows that finite width stripes very much resembling the experimental data can be obtained by more detailed simulations generalizing the latter model \cite{supplementary}.

Various physical mechanisms can influence the conductance along several steps of the evolution of the junction, and thus lead to a correlated shifting of the plateaus. These can be the trace to trace variations in the crystallographic orientations in the vicinity of the contact, changes in the contact geometries, or scattering on extended defects \cite{Sorensen98} being pinned to the contact region.

\paragraph{Discreteness of the minimal cross section.---}
Our results show some analogy to previous studies on work hardened Au and Al samples \cite{Yanson_Auworkhard,Yanson_Alworkhard}, where numerous peaks were
observed in the conductance histogram reflecting the quantized number of atoms in the MCS of the junction. In Au, where symmetric $s$ orbitals contribute to the conduction, the conductance is mainly  proportional to the MCS area \cite{Yanson_Auworkhard}, which explains that due to the discreteness of the MCS the conductance of the junction also prefers some discrete values.
In Al the observation of similar features is already somewhat surprising due to the role of $p$ orbitals in the conduction \cite{Yanson_Alworkhard}. In transition metals already up to six conductance channels can contribute to the conductance through a single atom \cite{Agrait_review,Jacob,Pauly_PtNi}, from which the conduction through low-symmetric $d$ orbitals is extremely sensitive to the directions of the bonds between the atoms. Therefore, contacts with the same MCS but different geometry (e.g. different crystallographic orientations) have a broad variance of conductances \cite{Pauly_PtNi} leading to a strong smearing of the histogram. Our measurements on Ni, Fe and V junctions show that -- in spite of the large number of plateaus in the conductance traces -- the histogram is indeed smeared, and the peaks are absent at higher conductance values. However, the 2DCHs can resolve the correlated shifting of plateaus giving a unique opportunity to follow the statistically relevant evolutions of the conductance staircase: once we fix the starting conductance of a high conductance plateau, it very well determines the subsequent plateau conductances during the rupture, as demonstrated by the arrows in Fig.~\ref{fig2.fig}(b).
These well-defined sequences of plateau conductances, and the numerous peaks in the conditional histograms demonstrate a very regular, well ordered rupture, indicating a well-defined discrete sequence of atomic configurations during the narrowing of the nanowire.

In contrast to the equidistant peak positions in the histograms of work hardened Au \cite{Yanson_Auworkhard}, the distances between the peaks in the
conditional histograms for Ni increase with increasing conductance. In the 2DCH the average difference in conductance between neighboring plateaus -- defined by the horizontal (or vertical) distance of the first stripe from the diagonal [Fig.~\ref{fig2.fig}(b)] -- accordingly shows a smooth increase. The low $G$ value of the conductance jump, $\Delta G \approx 1.6-2 \,G_0$ is fully consistent with the removal of a single atom from the MCS in each step. At higher conductances $\Delta G$ increases up to $3.5 \,G_0$, which can still be consistent with removing a single atom with several open conductance channels, assuming that the average conductance per atom is increasing with increasing junction size. Indeed, for $d$ metals it may happen that low coordinated atoms on the circumference of the junction give smaller contribution to the conductance than fully coordinated atoms in the contact center due to the role of $d$ orbitals oriented with large angle to the junction axis. Based on these simplified considerations the atom by atom narrowing of the junction is a possible explanation for the experimental observations. A regular, locally crystalline packing of the electrodes with slips of the lattice planes by atomic distances could also account for the observed regular rupture. However, an unambiguous microscopic explanation of the observed regular narrowing can only be given in comparison with detailed density functional theory calculations.

Contrary to Ni junctions our measurements on Au did not show any stripes in the 2DCH [Fig.~\ref{fig1.fig}(d)].
There is a significant difference in the mechanical behavior of two metals, which is clearly demonstrated by the traces presented in Figs.\ref{fig1.fig}(a) and \ref{fig2.fig}(c). In Ni junctions the conductance traces are very well ordered with well-defined conductance jumps [Fig.\ref{fig2.fig}(c)], whereas in Au -- though several plateaus are observed -- the conductance jumps shows a rather broad variety of amplitudes [Fig.\ref{fig1.fig}(a)]. This demonstrates, that -- although the numerous peaks in the histogram of Au reflect the discreteness of the minimal cross section -- the conductance histogram cannot tell how the MCS changes during a single conductance step. According to our measurements, in Au the changes of the atomic configurations are not regular, very frequently either multiple atoms are removed from the MCS, or just smaller rearrangements appear with $\Delta G<1\,$G$_0$, which smears the structures in the 2DCH. In Ni, however, the amplitude of the conductance jump is a very well-defined function of the junction size, as shown by the robust stripes in the 2DCH. This points to a very well ordered mechanical evolution of the junction, and indicates that the junction follows a well-defined sequence of atomic configurations. To check the relevancy of this unique behavior, we have performed measurements on a broad variety of metals \cite{Makk}. We have found that this regular atomic narrowing is absent in the majority of the metals, including the noble metals Au, Ag, Cu; the $p$ metals Al, Sn, Pb; and transition metals Ta, Nb. In Pt and Pd junctions we have detected some signs of regular narrowing, but it was less pronounced than in Ni, Fe and V.

\paragraph{Conclusions.---}
We have demonstrated, that several features of atomic-scale contact formation dynamics -- being hidden in traditional conductance histograms -- can be resolved by the 2D cross-correlation analysis of the conductance traces.
Applying this method we were able to identify a very regular atomic narrowing during the rupture of Ni, Fe and V nanowires, which is absent in the majority of the metals.

\paragraph{Acknowledgements.---}
The authors are grateful to  J.\ J. Palacios, F. Pauly, G. Sch\"on and J. van Ruitenbeek for
fruitful discussions. This work has been supported by the Hungarian research funds OTKA
K76010, NK72916, NNF78842, TAMOP-4.2.1/B-09/1/KMR-2010-0002 and the Polish grant for science in years
  2010-2013 as a research project. A.H.\ \& S.C\ were supported by a Bolyai
J\'anos Scholarship.

\end{document}